# Invariant recognition drives neural representations of action sequences

Andrea Tacchetti*, Leyla Isik* and Tomaso Poggio

Center for Brains Minds and Machines, MIT

*denotes equal contribution

## Abstract

*Recognizing the actions of others from visual stimuli is a crucial aspect of human visual perception that allows individuals to respond to social cues. Humans are able to identify similar behaviors and discriminate between distinct actions despite transformations, like changes in viewpoint or actor, that substantially alter the visual appearance of a scene. This ability to generalize across complex transformations is a hallmark of human visual intelligence. Advances in understanding motion perception at the neural level have not always translated in precise accounts of the computational principles underlying what representation our visual cortex evolved or learned to compute. Here we test the hypothesis that invariant action discrimination might fill this gap. Recently, the study of artificial systems for static object perception has produced models, Convolutional Neural Networks (CNNs), that achieve human level performance in complex discriminative tasks. Within this class of models, architectures that better support invariant object recognition also produce image representations that match those implied by human and primate neural data. However, whether these models produce representations of action sequences that support recognition across complex transformations and closely follow neural representations remains unknown. Here we show that spatiotemporal CNNs appropriately categorize video stimuli into actions, and that deliberate model modifications that improve performance on an invariant action recognition task lead to data representations that better match human neural recordings. Our results support our hypothesis that performance on invariant discrimination dictates the neural representations of actions computed by human visual cortex. Moreover, these results broaden the scope of the invariant recognition framework for understanding visual intelligence from perception of inanimate objects and faces in static images to the study of human perception of action sequences.*

## Author Summary

Recognizing the actions of others from video sequences across changes in viewpoint, actor, gait or illumination is a hallmark of human visual intelligence. A large number of studies have highlighted which areas in the human brain are involved in the processing of biological motion, and many more have described how single neurons behave in response to videos of human actions. However, little is known about the computational necessities that shaped these neural mechanisms either through evolution or experience. In this paper, we test the hypothesis that this computational goal is the discrimination of complex video stimuli according to their action content. We show that, within the class of Spatiotemporal Convolutional Neural Networks (ST-CNN), deliberate model modifications leading to representations of videos that better support robust action discrimination, also produce representations that better match human neural data. Because modifying a naïve model to increase its performance in invariant recognition resulted in a representation similar to human neural data, these results suggest that, similarly to what is known for object recognition, a process of performance optimization for invariant discrimination, constrained to a hierarchical ST-CNN-like architecture, shaped the neural mechanisms underlying our ability to perceive the actions of others.

## Introduction

Humans' ability to recognize the actions of others is a crucial aspect of visual perception. Remarkably, the accuracy with which we can finely discern what others are doing is largely unaffected by transformations that, while substantially changing the visual appearance of a given scene, do not change the semantics of what we observe (e.g. a change in viewpoint). Recognizing actions, the middle ground between action primitives and activities [1], across these transformations is a hallmark of human visual intelligence which has proven elusive to replicate in artificial systems. Because of this, invariance to transformations that are orthogonal to a learning task has been the subject of extensive theoretical and empirical investigation in both artificial and biological perception and recognition [2,3].

Recent studies have described the neural processing mechanisms underlying our ability to recognize the action of others from visual stimuli. Specific brain areas have been implicated in the processing of biological motion and the perception of actions. For example, in humans and other primates, the Superior Temporal Sulcus, and particularly its posterior portion, is believed to participate in the processing of biological motion and actions [4–11]. In addition to studying entire brain regions that engage during action recognition, a number of studies have characterized the responses of single neurons. The preferred stimuli of neurons in visual areas V1 and MT are well approximated by moving edge-detection filters and energy-based pooling mechanisms [12,13]. Neurons in the STS region of macaque monkeys respond selectively to actions, are invariant to changes in actors and viewpoint [14] and their tuning curves are well modeled by simple snippet-matching models [15].

More broadly, the neural computations underlying action recognition in visual cortex are organized as a hierarchical succession of spatiotemporal feature detectors of increasing size and complexity [16,17]. Despite precise characterization of the organizational, regional and single-unit data involved in the processing of biological motion, little information is known about what computational tasks might be relevant to explaining and recapitulating its computations.

Fueled by advances in computer vision methods for object and scene categorization, recent studies have made progress towards linking computational outcomes to neural signals through quantitatively accurate models of single neurons in Inferior Temporal cortex (IT) [18]. In addition, these studies have highlighted a correlation between performance optimization on discriminative object recognition tasks and the accuracy of neural predictions both at the single recording site and neural representation level [18–21].
However, these results, have not been extended to action perception and dynamic stimuli. What specific computational goals relate to the biological structure that underlies our ability to recognize the actions of others, and in particular to the representations of action sequences human visual cortex learned, or evolved to compute, remains unknown. Here we test our hypothesis that invariant recognition might fill this gap.

We use artificial systems for action recognition and compare their data representations to human Magnetoencephalography (MEG) recordings [22]. We show that, within the Spatiotemporal Convolutional Neural Networks model class [16,17,23,24], deliberate modifications that result in better performing models on invariant action recognition tasks, also lead to empirical dissimilarity matrices that better match those obtained by human neural recordings. Our results suggest that performance optimization on discriminative tasks, especially those that require generalization across complex transformations, alongside the constraints imposed by the hierarchical organization of motion processing in visual cortex, determined the representation of action sequences computed by visual cortex. Moreover, by highlighting the role of robustness to nuisances that are orthogonal to the discrimination task, our results extend the scope of invariant recognition as a computational framework for understanding human visual intelligence to the study of action recognition from video sequences.

## Results

### Action discrimination with Spatiotemporal Convolutional representations

We filmed a video dataset showing five actors, performing five actions (drink, eat, jump, run and walk) at five different viewpoints **[Figure 1]**. We then developed four variants of feedforward hierarchical models of visual cortex and used them to extract feature representations of videos showing two different viewpoints, frontal and side. Subsequently, we trained a machine learning classifier to discriminate video sequences into different action classes based on the model output. We then evaluated the classifier accuracy in predicting the action depicted in new, unseen videos.

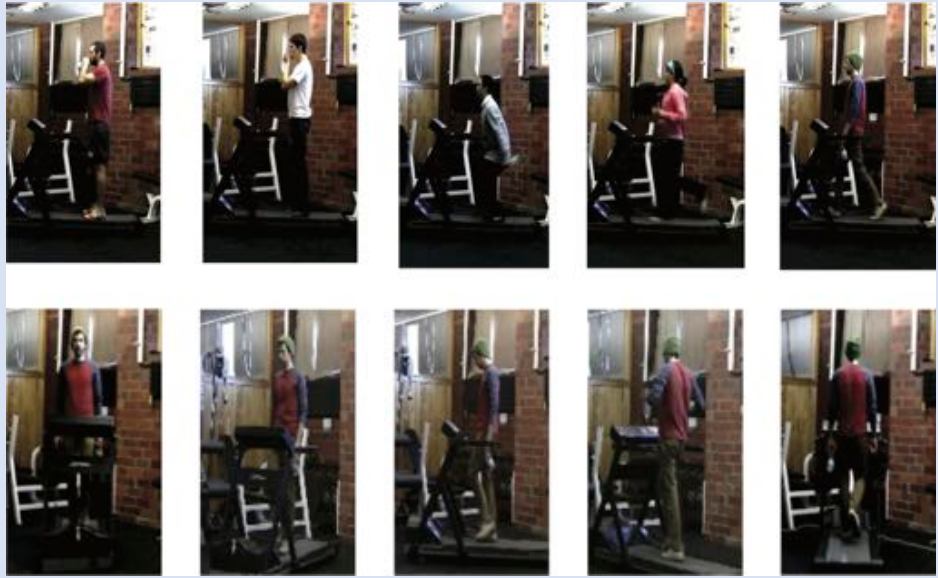

**Figure 1: Action recognition stimulus set**

Sample frames from action recognition dataset consisting of 2s video clips depicting five actors performing five actions (top row: drink, eat, jump, run and walk). Actions were recorded at five different viewpoints (bottom row: 0-frontal, 45, 90-side, 135 and 180 degrees with respect to the normal to the focal plane), they were all performed on a treadmill and actors held a water bottle and an apple in their hand regardless of the action they performed in order to minimize low-level object/action confounds. Actors were centered in the frame and the background was held constant regardless of viewpoint.

The four models we developed to extract video representations were instances of Spatiotemporal Convolutional Neural Networks (ST-CNNs), currently the best performing artificial perception systems for action recognition[23]. These architectures are direct extensions of the Convolutional Neural Networks used to recognize objects or faces in static images [25,26], to input stimuli that extend both in space and time. ST-CNNs are hierarchical models that build selectivity to specific stimuli through template matching operations and robustness to transformations through pooling operations **[Figure 2]**. Qualitatively, Spatiotemporal Convolutional Neural Networks detect the presence of a certain video segment (a template) in their input stimulus; detections for various templates are then aggregated, following a hierarchical architecture, to construct video representations. Nuisances that should not be reflected in the model output, like changes in position, are discarded through a pooling mechanism [27].

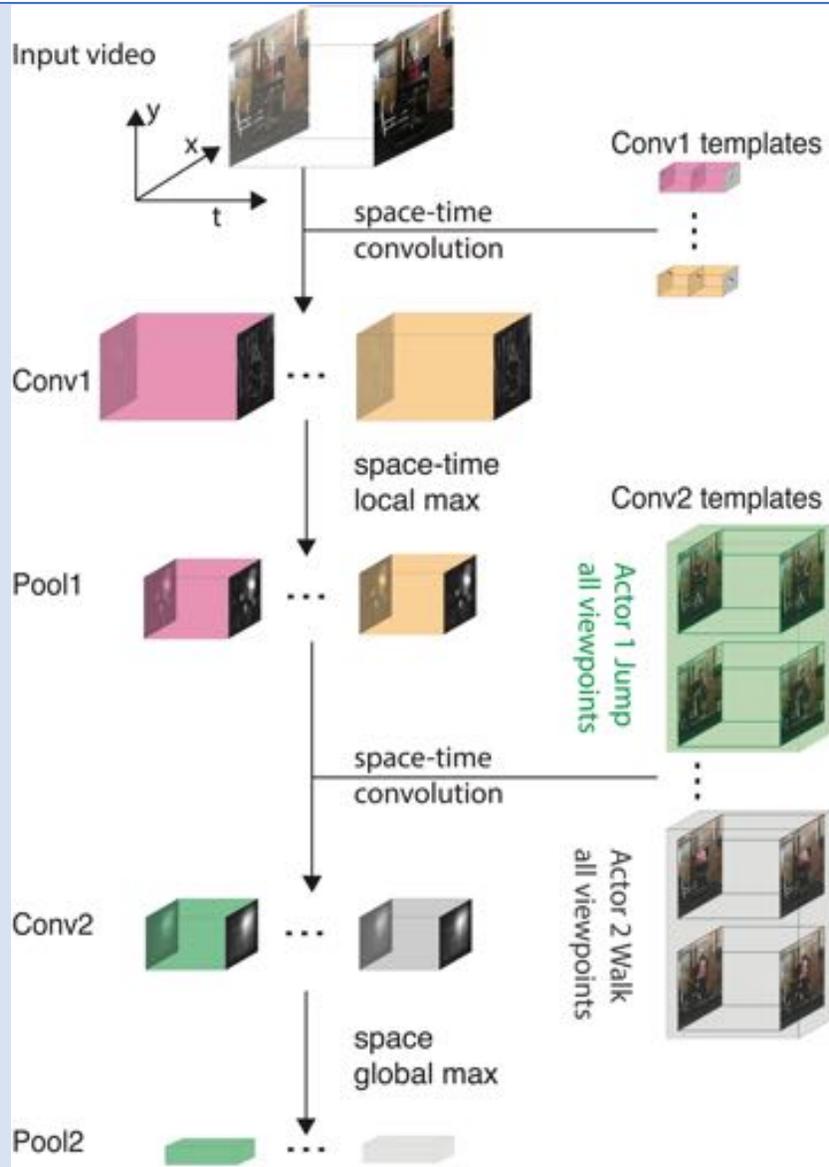

Figure 2: Spatiotemporal Convolutional Neural Networks

Schematic overview of the class of models we used: Spatiotemporal Convolutional Neural Networks (ST-CNNs). ST-CNNs are hierarchical feature extraction architectures. Input videos go through layers of computation and the output of each layer serves as input to the next layer. The output of the last layer constitutes the video representation used in downstream tasks. The models we considered consisted of two convolutional-pooling layers' pairs, denoted as Conv1, Pool1, Conv2 and Pool2. Convolutional layers performed template matching with a shared set of templates at all positions in space and time (spatiotemporal convolution), and pooling layers increased robustness through max-pooling operations. Convolutional layers' templates can be either fixed a priori, sampled or learned. In this example, templates in the first layer Conv1 are fixed and depict moving Gabor-like receptive fields, while templates in the second simple layer Conv2 are sampled from a set of videos containing actions and filmed at different viewpoints.

We considered a basic, purely convolutional model, and subsequently introduced modifications to its pooling mechanism and template learning rule to improve performance on invariant action recognition [26].

The first, purely convolutional model, consisted of convolutional layers with fixed templates, interleaved by pooling layers that computed max-operations across contiguous regions of space. In particular, templates in the first convolutional layer, contained moving Gabor filters while templates in the second convolutional layer were sampled from a set of action sequences collected at various viewpoints. The second, Unstructured Pooling model, allowed pooling layer units to span random sets of templates as well as contiguous space regions **[Figure 3b]**. The third, Structured Pooling model, allowed pooling over contiguous regions of space as well as across templates depicting the same action at various viewpoints. The 3D orientation of each template was discarded through this pooling mechanism, similarly to how position in space is discarded in traditional CNNs **[Figure 3a]** [2,28]. The fourth and final model employed backpropagation, a gradient based optimization method, to learn convolutional layers' templates by iteratively maximizing performance on an action recognition task [26]. We used each model to extract feature representations of action sequences at two different viewpoints, frontal and side. We then sought to evaluate the four models based on how well they could support discrimination between the five actions in our video dataset. We trained a machine learning classifier to discriminate stimuli based on action and later assessed classification accuracy on new, unseen videos. In Experiment 1, we trained and tested the classifier using model features computed from videos captured at the same viewpoint. In Experiment 2, we trained and tested the classifier using model features computed from videos at mismatching viewpoints (e.g. if the classifier was trained using videos captured at the frontal viewpoint, then testing would be conducted using videos at the side viewpoint).

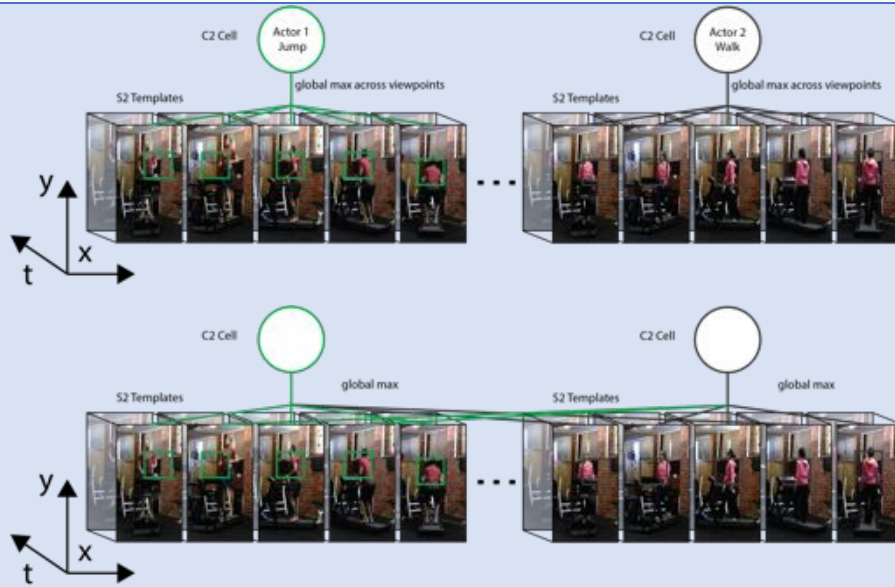

Figure 3: Structured and Unstructured Pooling

We introduced modifications to the basic ST-CNN to increase robustness to changes in 3D-viewpoint. Qualitatively Spatiotemporal Convolutional Neural Networks detect the presence of a certain video segment (a template) in their input stimulus. The 3D orientation of this template is discarded by the pooling mechanism in our structured pooling model, analogous to how position in space is discarded in a traditional CNN. a) In models with Structured Pooling, the template set for Conv2 layer cells was sampled from a set of videos containing four actors performing five actions at five different viewpoints (see Methods). All templates sampled from videos of a specific actor and performing a specific action were pooled together by one Pool2 layer unit. b) Models employing Unstructured Pooling allowed Pool2 cells to pool over the entire spatial extent of their input as well as across channels. These models used the exact same templates employed by models relying on Structured Pooling and matched these models in the number of templates wired to a pooling unit. However, the assignment of templates to pooling was randomized (uniform without replacement) and did not reflect any semantic structure.

Experiment 1: Action discrimination - viewpoint match condition

In Experiment 1, we trained and tested the action classifier using feature representations of videos acquired at the same viewpoint, and therefore did not investigate robustness to changes in viewpoint. All models produced representations that successfully classified videos based on the action they depicted **[Figure 4]**. We observed a significant difference in performance between model 4, the end-to-end trainable model, and fixed template models 1, 2 and 3 (see Methods Section). However, the task considered in Experiment 1 was not sufficient to rank the four types of ST-CNN models.

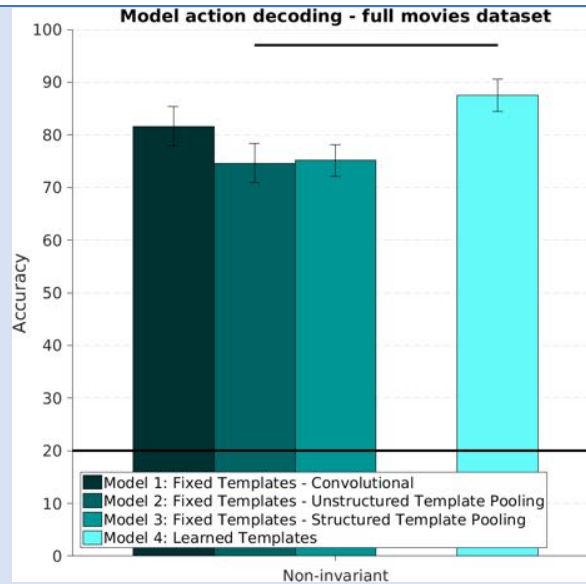

Figure 4: Action recognition: viewpoint match condition

We trained a supervised machine learning classifier to discriminate videos based on their action content by using the feature representation computed by each of the Spatiotemporal Convolutional Neural Network models we considered. This figure shows the prediction accuracy of a machine learning classifier trained and tested using videos recorded at the same viewpoint. The classifier was trained on videos depicting four actors performing five actions at either the frontal or side view. The machine learning classifier accuracy was then assessed using new, unseen videos of a new, unseen actor performing those same five actions. No generalization across changes in 3D viewpoints was required of the feature extraction and classification system. Here we report the mean and standard error of the classification accuracy over the five possible choices of test actor. Models with learned templates outperform models with fixed templates significantly on this task. Chance is 1/5 and is indicated by a horizontal line. Horizontal lines at the top indicate significant difference between two conditions ($p < 0.05$) based on group ANOVA or Bonferroni corrected paired t-test (see Methods section).

Experiment 2: Action discrimination - viewpoint mismatch condition

The four ST-CNN models we developed were designed to have varying degrees of tolerance to changes in viewpoint. In Experiment 2, we investigated how well these model representations could support learning to discriminate video sequences based on their action content, across changes in viewpoint. The general experimental procedure was identical to the one outlined for Experiment 1, except we used features extracted from videos acquired at mismatching viewpoints for training and testing (e.g., a classifier trained using videos captured at the frontal viewpoint, would be tested on videos at the side viewpoint). All the models we considered produced representations that were, at least to a minimal degree, useful to discriminate actions invariantly to changes in viewpoint **[Figure 5]**. Unlike what we observed in Experiment 1, it was possible to rank the models we considered based on performance on this task. This was expected, since the various models were designed to exhibit various degrees of robustness to changes in viewpoint (see Methods Section). The end-to-end trainable models (model 4) performed better than models 1,2 and 3, which used fixed templates, on this task. Within the fixed templates models group, as expected, models that employed a Structured Channel Pooling mechanism to increase robustness performed best [29].

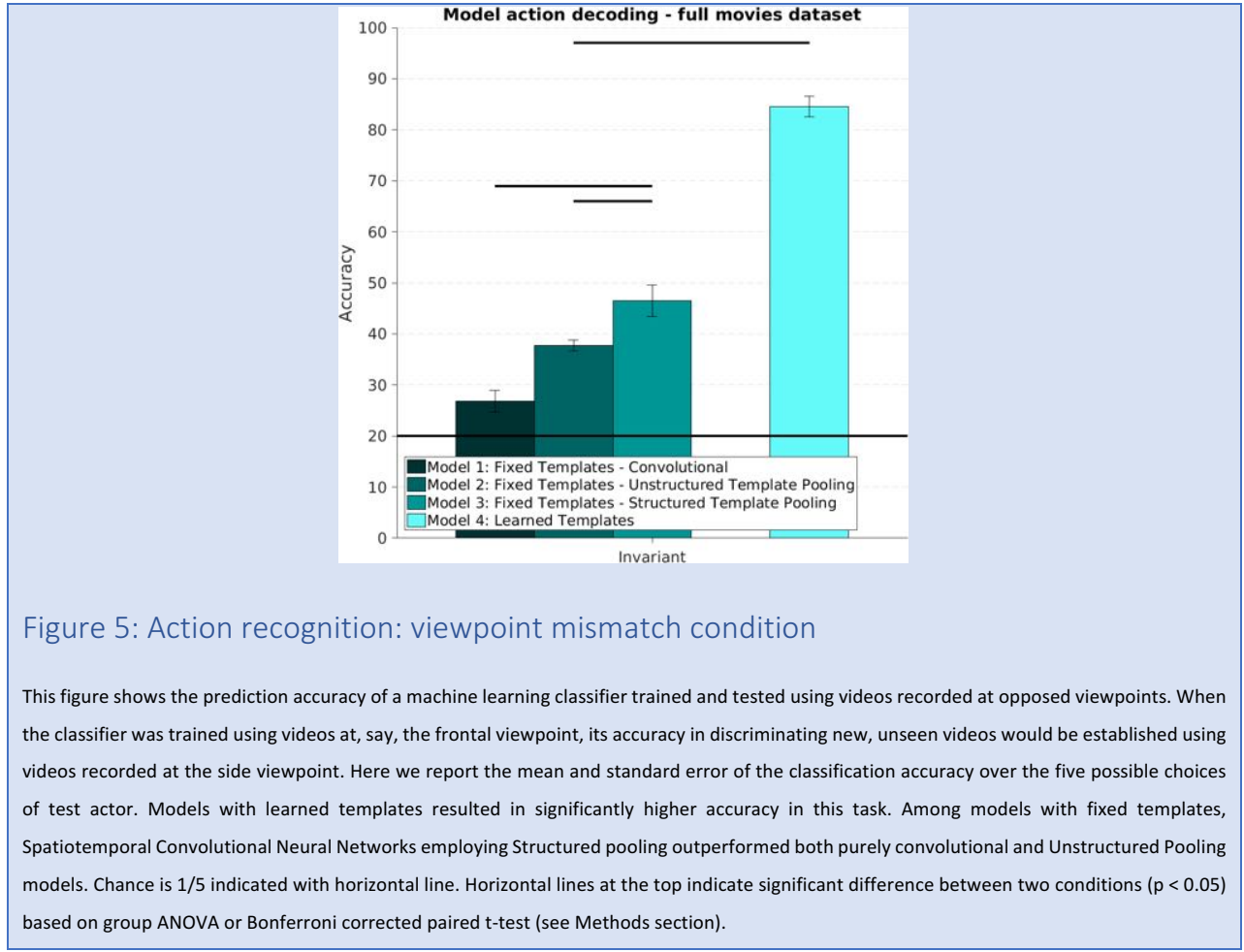

Figure 5: Action recognition: viewpoint mismatch condition

This figure shows the prediction accuracy of a machine learning classifier trained and tested using videos recorded at opposed viewpoints. When the classifier was trained using videos at, say, the frontal viewpoint, its accuracy in discriminating new, unseen videos would be established using videos recorded at the side viewpoint. Here we report the mean and standard error of the classification accuracy over the five possible choices of test actor. Models with learned templates resulted in significantly higher accuracy in this task. Among models with fixed templates, Spatiotemporal Convolutional Neural Networks employing Structured pooling outperformed both purely convolutional and Unstructured Pooling models. Chance is 1/5 indicated with horizontal line. Horizontal lines at the top indicate significant difference between two conditions (p < 0.05) based on group ANOVA or Bonferroni corrected paired t-test (see Methods section).

## Comparison of model representations and neural recordings

We used Representational Similarity Analysis (RSA) to assess how well each model feature representation matched the human neural data. RSA produces a measure of agreement between artificial models and brain recordings based on the correlation between empirical dissimilarity matrices constructed using either the model representation of a set of stimuli, or recordings of the neural responses these stimuli elicit **[Figure 6]** [21]. We used video feature representations extracted by each model from a set of new, unseen stimuli to construct model dissimilarity matrices and Magnetoencephalograpy (MEG) sensor recordings of the neural activity elicited by those same stimuli to compute a neural dissimilarity matrix [22]. Finally, we constructed a dissimilarity matrix using an action categorical oracle. In this case, the dissimilarity between videos of the same action was zero and the distance across actions was one.

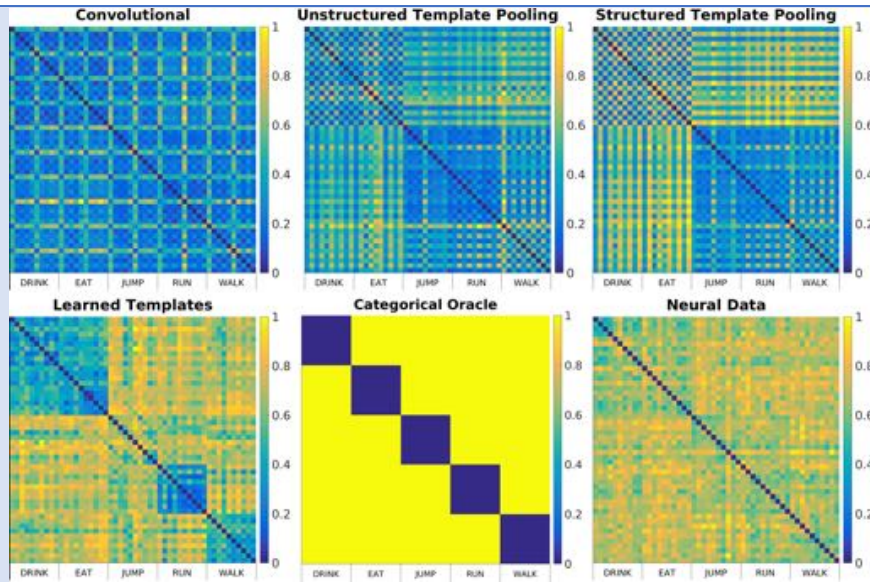

Figure 6: Feature representation empirical dissimilarity matrices

We used feature representations, extracted with the four Spatiotemporal Convolutional Neural Network models, from 50 videos depicting five actors performing five actions at two different viewpoints, frontal and side. Moreover, we obtained Magnetoencephalography (MEG) recordings of human subjects' brain activity while they were watching these same videos, and used these recordings as a proxy for the neural representation of these videos. These videos were not used to construct or learn any of the models. For each of the six representations of each video (four artificial models, a categorical oracle and one neural recordings) we constructed an empirical dissimilarity matrix using linear correlation and normalized it between 0 and 1. Empirical dissimilarity matrices on the same set of stimuli constructed with video representations from a) Model 1: Purely Convolutional model, b) Model 2: Unstructured pooling model, c) Model 3: Structured pooling model d) Model 4: Learned templates model e) Categorical oracle and f) Magnetoencephalography brain recordings.

We observed that end-to-end trainable model (model 4) produced dissimilarity structures that better agreed with those constructed from neural data than models with fixed templates **[Figure 7]**. Within models with fixed templates, model 3, constructed using a Structured Pooling mechanism to build invariance to changes in viewpoint, produced representations that agree better with the neural data than models employing Unstructured Pooling (model 2) and purely convolutional models (model 1).

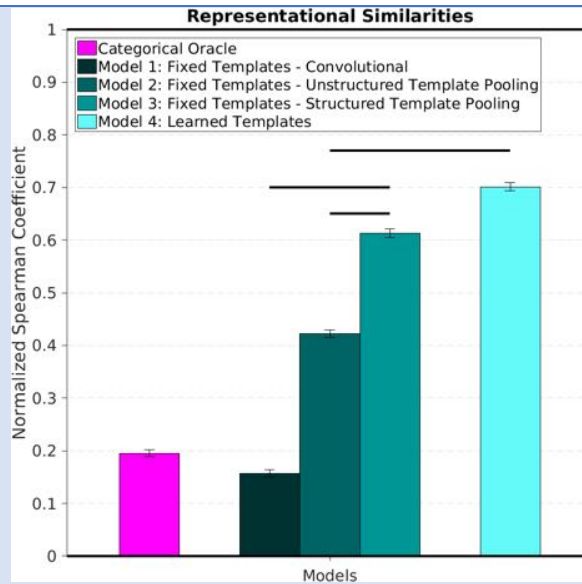

Figure 7: Representational Similarity Analysis between Model representations and Human Neural data

We computed the Spearman Correlation Coefficient (SCC) between the lower triangular portion of the dissimilarity matrix constructed with each of the artificial models we considered and the dissimilarity matrix constructed with neural data (shown and described in Figure 6). We assessed the uncertainty of this measure by resampling the rows and columns of the matrices we constructed. The noise ceiling was assessed by computing the SCC between each individual human subjects' dissimilarity matrix and the average dissimilarity matrix over the rest of the subjects. The noise floor was computed by assessing the SCC between the lower portion of the dissimilarity matrix constructed using each of the model representation and a scrambled version of the neural dissimilarity matrix. The score reported here is normalized so that the noise ceiling is 1 and the noise floor is 0. Models with learned templates agree with the neural data significantly better than models with fixed templates. Among these, models with Structured Pooling outperform both purely Convolutional and Unstructured models. Horizontal lines at the top indicate significant difference between two conditions ($p < 0.05$) based on group ANOVA or Bonferroni corrected paired t-test (see Methods section).

## Discussion

We have shown that, within the Spatiotemporal Convolutional Neural Networks model class, and across a deliberate set of model modifications, feature representations that are more useful to discriminate actions in video sequences, in a manner that is robust to changes in viewpoint, also produce empirical dissimilarity structures that are more similar to those constructed using human neural data. Discrimination performance on a simpler task, that does not require generalization across complex transformations, was not sufficient to fully rank the model representations. These results support our hypothesis that performance on invariant discriminative tasks shaped the neural representations of actions that are computed by our visual cortex. Moreover, dissimilarity matrices constructed with ST-CNNs representations match those build with neural data better than a purely categorical dissimilarity matrix. This highlights the importance of both the computational task and the architectural constraints, described in previous accounts of the neural processing of action and motions, to build quantitatively accurate models of neural data representations [30]. Our findings are in agreement with what has been reported for the perception of objects from static images, both at the single recording site and at the whole brain level [18–20], and identify a computational task that explains and recapitulates the representations of human action in visual cortex.

We developed the four ST-CNN models using deliberate modifications to improve the models' feature representations to invariant action recognition. In so doing, we verified that structured pooling architectures and memory based learning (model 3), as previously described and theoretically motivated [2,3], can be applied to build representations of video sequences that support recognition invariant to complex, non-affine transformations. However, empirically, we found that learning model templates using gradient based methods and a fully supervised action recognition task (model 4), led to better results, both in terms of classification accuracy and agreement with neural recordings [20].

A limitation of the methods we employed is that the extent of the match between a model representation and the neural data is appraised solely based on the correlation between the empirical dissimilarity structures constructed with neural recordings and model representations. This relatively abstract comparison provides no guidance in establishing a one-to-one mapping between model units and brain regions or sub-regions and therefore cannot exclude models on the basis of biological implausibility [19]. In this work, we mitigated this limitation by constraining the model class to reflect previous accounts of neural computational units and mechanisms that are involved in the perception of motion [12,13,17,31,32].

Furthermore, the class of models we developed in our experiments is purely feedforward, however, the neural recordings we selected were acquired 470ms after stimulus onset. This late in the visual processing, it is likely that feedback signals are among the energy sources captured by the recordings. These signals are not accounted for in our models. We provide evidence that adding a feedback mechanism, through recursion, does not improve recognition performance nor correlation with the neural data **[Supplementary Figure 1]**. We cannot, however, exclude that this is due to the stimuli and discrimination task we designed, which only considered pre-segmented, relatively short action sequences.

Recognizing the actions of others from complex visual stimuli is a crucial aspect of human perception. We investigated the relevance of invariant action discrimination to improving model representations' agreement with neural recordings and showed that it is one of the computational principles shaping the representation of human action sequences human visual cortex evolved, or learned to compute. Our deliberate approach to model design underlined the relevance of both supervised, gradient based, performance optimization methods and memory based, structured pooling methods to the modeling of neural data representations. If and how primate visual cortex could implement gradient based optimization or acquire the necessary supervision remains, despite recent efforts, an unsettled matter [33–35]. Memory-based learning and structured pooling have been investigated extensively as a biologically plausible learning algorithms [2,36–38]. Irrespective of the precise biological mechanisms that could carry out performance optimization on invariant discriminative tasks, computational studies point to its relevance to understanding neural representations of visual scenes [18–20]. Recognizing the semantic category of visual stimuli across photometric, geometric or more complex changes, in very low sample regimes is a hallmark of human visual intelligence. By building data representations that support this kind of robust recognition, we have shown here, one obtains empirical dissimilarity structures that match those constructed using human neural data. In the wider context of the study of perception, our results strengthen the claim that the computational goal of human visual cortex is to support invariant recognition by broadening it to the study of action perception.

## Materials and methods

### Action Recognition Dataset

We collected a dataset of five actors performing five actions (drink, eat, jump, run and walk) on a treadmill at five different viewpoints (0, 45, 90, 135 and 180 degrees between the line across the center of the treadmill and the line normal to the focal plane of the video-camera). We rotated the treadmill rather than the camera to keep the background constant across changes in viewpoint **[Figure 1]**. The actors were instructed to hold an apple and a bottle in their hand regardless of the action they were performing, so that objects and background would not differ between actions. Each action/actor/view was filmed for at least 52s. Subsequently the original videos were cut into 26 clips, each 2s long resulting in a dataset of 3,250 video clips. Video clips started at random points in the action cycle (for example a jump might start mid-air or before the actor's feet left the ground) and each 2s clip contained a full action cycle. The authors manually identified one single spatial bounding box that contained the entire body of each actor and cropped all videos according to this bounding box.

### Recognizing actions with spatiotemporal convolutional representations

#### General Experimental Procedure

Experiment 1 and Experiment 2 were designed to quantify the amount of action information extracted from video sequences by four computational models of primate visual cortex. In Experiment 1, we tested basic action recognition. In Experiment 2, in particular, we further quantified whether this action information could support action recognition robustly to changes in viewpoint. The motivating idea behind our design is that, if a machine learning classifier is able to discriminate unseen video sequences based on their action content, using the output of a computational model, then this model representation contains some action information.

Moreover, if the classifier is able to discriminate videos based on action at new, unseen viewpoints, using model outputs then it must be that these model representations not only carry action information, but that changes in viewpoint are not reflected in the model output. This procedure is analogous to neural decoding techniques with the important difference that the output of an artificial model is used in lieu of brain recordings [39,40].

The general experimental procedure is as follows: we constructed feedforward hierarchical spatiotemporal convolutional models and used them to extract feature representations of a number of video sequences. We then trained a machine learning classifier to predict the action label of a video sequence based on its feature representation. Finally, we quantified the performance of the classifier, by measuring prediction accuracy on a set of new, unseen videos.

The procedure outlined above was performed using three separate subsets of the action recognition dataset described in the previous section. In particular, constructing spatiotemporal convolutional model requires access to video sequences depicting actions to sample or learn convolutional layers' templates. The subset of video sequences used to learn or sample templates was called **"embedding set"**. Training and testing the classifier required extracting model responses from a number of video sequences; these sequences were organized in two subsets: **"training set"** and **"test set"**. There was never any overlap between the **"test set"** and the union of **"training set"** and **"embedding set"**.

Experiment 1

The purpose of Experiment 1 was to assess how well the data representations produced by each of the four models, supported a non-invariant action recognition task. In particular, the **embedding set** used to sample or learn templates contained videos showing all five actions at all five viewpoints performed by four of the five actors. The **training set** was a subset of the embedding set, and contained only videos at either the frontal viewpoint or the side viewpoint. Lastly the **test set** contained videos of all five actions, performed by the fifth left-out actor and performed at either the frontal or side viewpoint. We obtained five different splits by choosing each of the five actors exactly once for test. After the templates had either been learned or sampled we used each model to extract representations of the **train** and **test sets** videos. We averaged the performance over the two possible choices of training viewpoint, frontal or side. We report the mean and standard error of the classification accuracy across the five possible choices of the test actor.

Experiment 2

Experiment 2 was designed to assess the performance of each model in producing data representations that were useful to classify videos according to their action content, when a generalization across changes in viewpoint was required. The experiment is identical to Experiment 1, and used the exact same models. However, when the **training set** contained videos recorded at the frontal viewpoint, the **test set** would contain videos at side viewpoint and vice-versa. We report the mean and standard deviation over the choice of the test actor of the average accuracy over the choice of training viewpoint.

Feedforward Spatiotemporal Convolutional Neural Networks

Feedforward Spatiotemporal Convolutional Neural Networks (ST-CNNs) are hierarchical models: input video sequences go through layers of computations and the output of each layer serves as input to the next layer **[Figure 2]**. These models are direct generalizations of models of the neural mechanisms that support recognizing objects in static images [27,41], to stimuli that extend in both space and time (i.e. video stimuli). Within each layer, single computational units process a portion of the input video sequence that is compact both in space and time. The outputs of each layer's units are then processed and aggregated by units in the subsequent layers to construct a final signature representation for the whole input video. The sequence of layers we adopted alternates layers of units which perform template matching (or convolutional layers), and layers of units which perform max pooling operations [15,17,23]. Units' receptive field sizes increases as the signal propagates through the hierarchy of layers.

All convolutional units within a layer share the same set of templates (filter bank) and output the dot-product between each filter and their input. Qualitatively, these models work by detecting the presence of a certain video segment (a template) in the input stimulus. The exact position in space and time of the detection is discarded by the pooling mechanism. The specific models we present here consist of two convolutional-pooling layers' pairs. The layers are denoted as Conv1, Pool1, Conv2 and Pool2 **[Figure 2]**. Convolutional layers are completely characterized by the size, content and stride of their units' receptive fields and pooling layers are completely characterized by the operation they perform (in the cases we considered, output the maximum value of their input) and their pooling regions (which can extend across space, time and filters).

Model 1: Purely convolutional models with sampled templates

The purely convolutional models with fixed and sampled templates we considered were implemented using the Cortical Network Simulator package [42].

The input videos were (128x76 pixel) x 60 frames; the model received the original input videos alongside two scaled-down versions of it (scaling of factors ½ and ¼ in each spatial dimension respectively).

The first layer, Conv1, consisted of convolutional units with 72 templates of size (7x7 pixel) x 3 frames, (9x9 pixel) x 4 frames and (11x11 pixel) x 5 frames. Convolution was carried out with a stride of 1 pixel (no spatial subsampling). Conv1 filters were obtained by letting Gabor-like receptive fields shift in space over frames (as described in previous studies describing the receptive fields of V1 and MT cells [12,13,31]). The full expression for each filter was as follows:

$$G(x, y, t, \theta, \rho, \sigma, \lambda, n) = f(t) \exp\left(-\frac{(x'(\theta, \rho, t)^2 + y'(\theta, \rho, t)^2)}{2\sigma^2}\right) \cos\left(\frac{2\pi y'}{\lambda}\right)$$

Where $x'(\theta, \rho, t)$ and $y'(\theta, \rho, t)$, are transformed coordinates that take into account a rotation by $\theta$ and a shift by $\rho t$ in the direction orthogonal to $\theta$. The Gabor filters we considered had a spatial aperture (in both spatial directions) of $\sigma = 0.6\,S$, with $S$ representing the spatial receptive field size and a wavelength $\lambda = \frac{\sqrt{2}}{2}\sigma$ [43]. Each filter had a preferred orientation $\theta$ chosen among 8 possible orientations (0, 45, 90, 135, 180, 225, 270, 315 degrees with respect to vertical). Each template was obtained by letting the Gabor-like receptive field just described, shift in the orthogonal direction to its preferred orientation (e.g. a vertical edge would move sideways) with a speed $\rho$ chosen from a linear grid of 3 points between 4/3 and 4 pixels per frame (the shift in the first frame of the template was chosen so that the mean of Gabor-like receptive field's envelop would be centered in the middle frame). Lastly, Conv1 templates had time modulation $f(t) = (kt)^2 e^{-kt^2}\left[\frac{1}{n!} - \frac{(kt)^2}{(n+2)!}\right]$ with $n = 3$ and $t = 0, \ldots, T$ with $T$ the temporal receptive field size [13,17].

The second layer, Pool1, performed max pooling operations on its input by simply finding and outputting the maximum value of each pooling region. Responses to each channel in the Conv1 filter bank was pooled independently and units pooled across regions of space: (4x4 units in space) x 1 unit in time with a stride of 2 units in space, and 1 unit in time, and two scale channels.

A second simple layer Conv2, followed Pool1. Templates in this case were sampled randomly from the Pool1 responses to videos in the **embedding set**. We used a model with 512 Conv2 units with sizes (9x9 units in space) x 3 units in time, (17x17 units in space) x 7 units in time and (25x25 units in space) x 11 units in time, and stride of 1 in all directions.

Finally, the Pool2 layer units performed max pooling. Pooling regions extended over the entire spatial input, one temporal unit, all remaining scales, and a single Conv2 channel.

Models 2 and 3: Structured and Unstructured Pooling models with sampled templates

Structured and Unstructured Pooling models (model 2 and 3, respectively) were constructed by modifying the Pool2 layer of the purely convolutional models. Specifically, in these models Pool2 units pooled over the entire spatial input, one temporal unit, all remaining scales and 8 or 9 Conv2 (512 channels and 60 Pool2 units) channels.

In the models employing a Structured Pooling mechanism, all templates sampled from videos of a particular actor performing a particular action, regardless of viewpoint were pooled together [**Figure 3b**]. Templates of different sizes and corresponding to different scale channels were pooled independently. This resulted in 6 Pool2 units per action/actor pair, one for each receptive-field-size/scale-channel pair. The intuition behind the Structured Pooling mechanism is that the resulting Pool2 units will respond strongly to the presence of a certain template (e.g. the torso of someone running) regardless of its 3D pose [2,28,29,44–48].

The models employing an Unstructured Pooling mechanism followed a similar pattern however, the wiring between simple and complex cells was random **[Figure 3a]**. The fixed templates models (model 1,2 and 3) employed the exact same set of templates (we sampled the templates from the embedding sets only once and used them in all three models) and differed only in their pooling mechanisms.

### Model 4: Learned template models

Models with learned templates were implemented using Torch packages. These models' templates were trained to recognize actions from videos in the embedding set using a Cross Entropy Loss function, full supervision and backpropagation [49]. The models' general architecture was similar to the one we used for models with structured and unstructured pooling. Specifically, during template learning we used two stacked Convolution-BatchNorm-MaxPooling-BatchNorm modules [50] followed by two Linear-ReLU-BatchNorm modules (ReLU units are half-rectifiers) and a final Log-Soft-Max layer. During feature extraction, the Linear and LogSoftMax layers were discarded. Input videos were resized to (128x76 pixel) x 60 frames, like in the fixed-template models. The first convolutional layer's filter bank comprised 72 filters of size (9x9 pixel) x 9 frames and convolution was applied with stride of 2 in all directions. The first max-pooling layer used pooling regions of size (4x4 units in space) x 1 unit in time and were applied with stride of 2 units in both spatial directions and 1 unit in time. The second convolutional layer's filter bank was made up of 60 templates of size (17x17 units in space) x 3 units in time, responses were computed with a stride of 2 units in time and 1 unit in all spatial directions. The second MaxPooling layer's units pooled over the full extent of both spatial dimensions, 1 unit in time and 5 channels. Lastly, the Linear layers had 256 and 128 units respectively and free bias terms. Model training was carried out using Stochastic Gradient Descent [49] and mini-batches of 10 videos.

### Machine learning classifier

We used the GURLS package [51] to train and test a Regularized Least Squares Gaussian-Kernel classifier using features extracted from the training and test set respectively and the corresponding action labels. The aperture of the Gaussian Kernel as well as the l2 regularization parameter were chosen with a Leave-One-Out cross-validation procedure on the training set. Accuracy was evaluated separately for each class and then averaged over classes.

### Significance testing: Model accuracy

We used a group one-way ANOVA to assess the significance of the difference in performance between all the fixed-template methods and the models with learned templates. We then used a paired-sample t-test with Bonferroni correction to assess the significance level of the difference between the performance of individual models. Difference were deemed significant p < 0.05.

## Quantifying agreement between model representations and neural recordings

### Neural recordings

The brain activity of 8 human participants with normal or corrected to normal vision was recorded with an Elekta Neuromag Triux Magnetoencephalography (MEG) scanner while they watched 50 videos (five actors, five actions, two viewpoints: front and side) acquired with the same procedure outlined above, but not included in the dataset used for model template sampling or training. The MEG recordings data was first presented in [22] (the reference also details all acquisition, preprocessing and decoding methods). In the original neural recording study MEG recordings were used to train a pattern classifier to discriminate video stimuli on the basis of the neural response they elicited. The performance of the pattern classifier was then assessed on a separate set of recordings from the same subjects. This train/test decoding procedure was repeated every 10ms and individually for each subject both in a non-invariant (train and test at the same viewpoint) and an invariant (train at one viewpoint and test at the different viewpoint) case. It was possible to discriminate videos based on their action content based on the neural response they elicited [22].

We used the filtered MEG recordings (all 306 sensors) elicited by each of the 50 videos mentioned above, averaged across subjects and averaged over a 100ms window centered around 470ms after stimulus onset as a proxy to the neural representation of the video (maximum accuracy for action decoding).

### Representational Similarity Analysis

We computed the pairwise correlation-based dissimilarity matrix for each of the model representations of the 50 videos that were shown to human subjects in the MEG. Likewise, we computed the empirical dissimilarity matrix computed using MEG neural recordings. We then performed 50 rounds of bootstrap, in each round we randomly sampled 30 videos out of the original 50 (corresponding to 30 rows and columns of the dissimilarity matrices). For each 30-videos sample, we assessed the level of agreement of the dissimilarity matrix induced by each model representation, with the one computed using neural data by calculating the Spearman Correlation Coefficient (SCC) between the lower triangular portions of the two matrices.

We computed an estimate for the noise ceiling in the neural data by repeating the bootstrap procedure outlined above to assess the level of agreement between an individual human subject and the average of the rest. We then selected the highest possible match score across subjects and across 100 rounds of bootstrap to serve as noise ceiling.

Similarly, we assessed a chance level for the Representational Similarity score by computing the match between each model and a scrambled version of the neural data matrix. We performed 100 rounds of bootstrap per model (reshuffling the neural dissimilarity matrix rows and columns each time) and selected the maximum score across rounds of bootstrap and models to serve as baseline score [21].

We normalized the SCC obtained by comparing each model representation to the neural recordings, by re-scaling them to fall between 0 (chance level) and 1 (noise ceiling). In this normalized scale, anything positive matches neural data better than chance with $p < 0.01$.

### Significance testing: Matching neural data

We used a one-way group ANOVA to assess the difference between the Spearman Correlation Coefficient (SCC) obtained using models that employed fixed templates and models with learned templates. Subsequently, we assessed the significance of the difference between the SCC of each model by performing a paired t-test between the samples obtained through the bootstrap procedure. We deemed differences to be significant when $p < 0.05$ (Bonferroni corrected).

# Acknowledgements

We would like to thank Georgios Evangelopoulos, Charles Frogner, Patrick Winston, Gabriel Kreiman, Martin Giese, Charles Jennings, Heuihan Jhuang, and Cheston Tan for their feedback on this work. This material is based upon work supported by the Center for Brains, Minds and Machines (CBMM), funded by NSF STC award CCF-1231216. We gratefully acknowledge the support of NVIDIA with the donation of the Tesla K40 GPU used for this research. We are grateful to the Martinos Imaging Center at MIT, where the neural recordings used in this work were acquired and to the McGovern Institute for Brain Research at MIT for supporting this research.

# Supplementary information

## Recurrent Neural Networks

We constructed a Recurrent version of our hierarchical model and assessed how well the resulting representation supports invariant and non-invariant action recognition as well as how closely it matches neural data, in order to try and account for these feedback mechanisms. The network architecture is identical to the Model with Learned Templates used for Experiment 1, Experiment 2 and Experiment 3 (see Methods section). The Linear layers, however, were replaced by a Recurrent structure with a hidden state of 128 units, resulting in an architecture similar to the one described in [52]. Classification accuracy on an invariant and a non-invariant action recognition tasks as well as quality of match between the representation and neural data are presented in **[Supplemental Figure 1]**. All procedures and data splits are identical to those described in the Methods section in the main text. Recurrent neural networks do not exceed the performance of simpler Feedforward models on the particular action recognition task presented in this work, however it is hard to draw definitive conclusions about their relevance to neural modeling, given the fact that the stimuli we used were relatively short and contained pre-segmented actions.

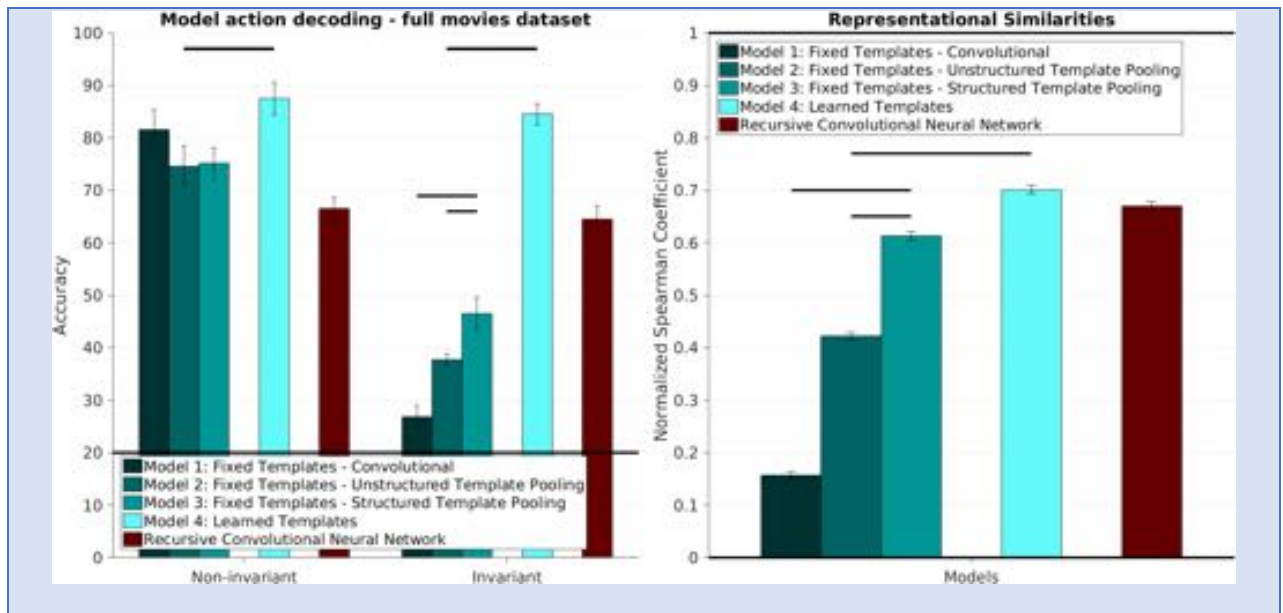

### Supplementary Figure 1

a) Classification accuracy, within and across changes in 3D viewpoint for a Recurrent Convolutional Neural Network. This architecture does not outperform a purely feedforward baseline. b) A Recurrent Convolutional Neural Network does not produce a dissimilarity structure that better agrees with the neural data than a purely feedforward baseline.

## RSA over time

For completeness, we report here the absolute value of the normalized Spearman Correlation Coefficient between the model with learned templates and the neural data. The time series is obtained by repeating the procedure outlined in the methods section using neural data from each time point (stimulus onset is at time 0) and time point used for figures in the main text corresponds to the vertical black line. Results are presented in **[Supplementary Figure 2]**.

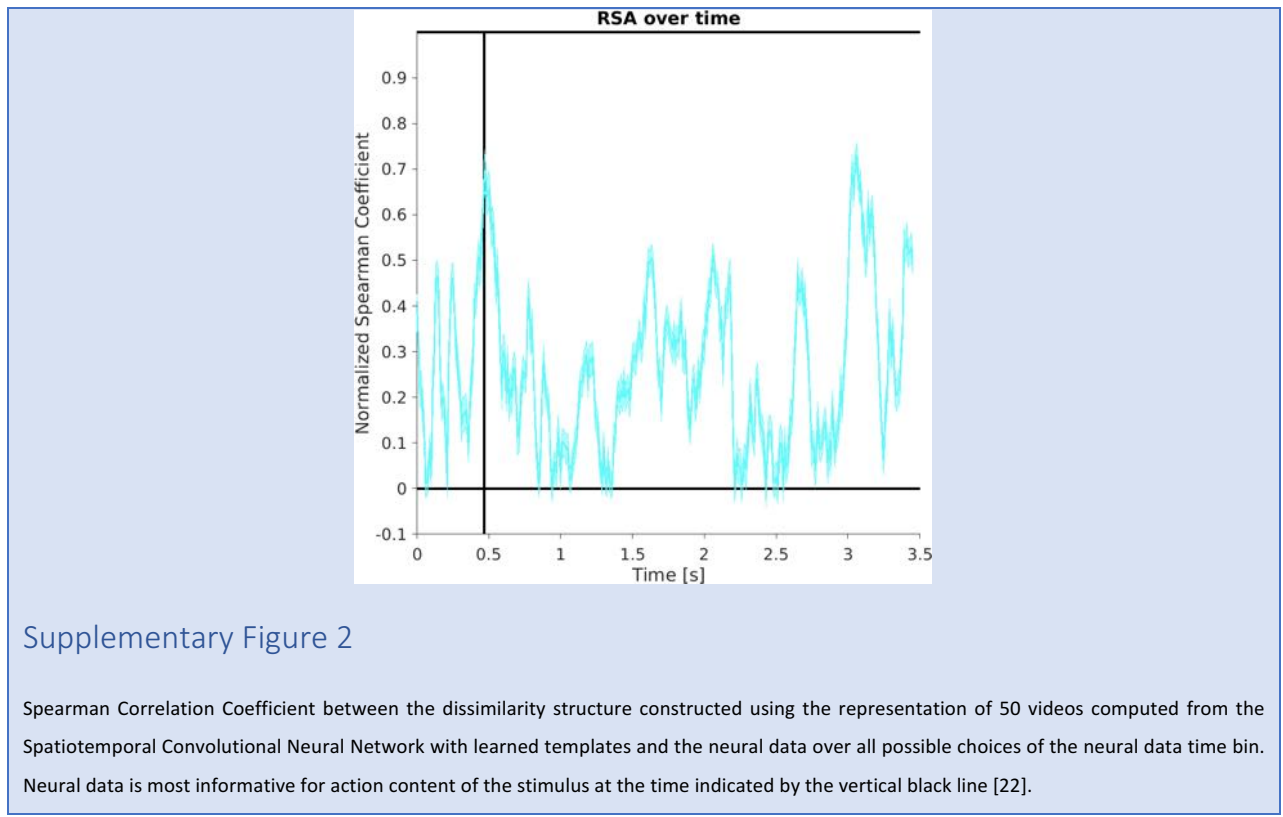

### Supplementary Figure 2

Spearman Correlation Coefficient between the dissimilarity structure constructed using the representation of 50 videos computed from the Spatiotemporal Convolutional Neural Network with learned templates and the neural data over all possible choices of the neural data time bin. Neural data is most informative for action content of the stimulus at the time indicated by the vertical black line [22].